\documentclass[preprint,12pt]{elsarticle}

\usepackage{amssymb}
\usepackage{color}
\usepackage{soul}
\usepackage[colorlinks=true]{hyperref}
\usepackage[]{cleveref}

\newcounter{bla}

\def\suspect3{{\tt SuSpect3}}
\def\suspect{{\tt SuSpect}}

\journal{Computer Physics Communications}

\newcommand{\quark}{\ensuremath{\mathrm{q}}}

\newcommand{\uQuark}{\ensuremath{\mathrm{u}}}
\newcommand{\cQuark}{\ensuremath{\mathrm{c}}}
\newcommand{\tQuark}{\ensuremath{\mathrm{t}}}

\newcommand{\dQuark}{\ensuremath{\mathrm{d}}}
\newcommand{\sQuark}{\ensuremath{\mathrm{s}}}
\newcommand{\bQuark}{\ensuremath{\mathrm{b}}}

\newcommand{\ele}{\ensuremath{\mathrm{e}}}

\newcommand{\Zzero}{\ensuremath{\mathrm{Z}^\circ}}

\newcommand{\gev}{\ensuremath{\mathrm{GeV}}}
\newcommand{\tev}{\ensuremath{\mathrm{TeV}}}

\newcommand{\MGUT}{\ensuremath{\mathrm{M}_{\mathrm{GUT}}}}
\newcommand{\MHIGH}{\ensuremath{\mathrm{M}_{\mathrm{High}}}}
\newcommand{\MMESS}{\ensuremath{\mathrm{M}_{\mathrm{MESS}}}}
\newcommand{\MEWSB}{\ensuremath{\mathrm{M}_{\mathrm{EWSB}}}}
\newcommand{\MMZ}{\ensuremath{\mathrm{M}_{\Zzero}}}

\newcommand{\tanb}{\ensuremath{\tan\beta}}

\newcommand{\mhalf}{\ensuremath{m_{1/2}}}
\newcommand{\Mone}{\ensuremath{\mathrm{M}_1}}
\newcommand{\Mtwo}{\ensuremath{\mathrm{M}_2}}
\newcommand{\Mthree}{\ensuremath{\mathrm{M}_3}}

\newcommand{\mzero}{\ensuremath{m_{\circ}}}

\newcommand{\Hu}{\ensuremath{\mathrm{H}_{\uQuark}}}
\newcommand{\Hd}{\ensuremath{\mathrm{H}_{\dQuark}}}
\newcommand{\mHu}{\ensuremath{\mathrm{M}_{\Hu}}}
\newcommand{\mHd}{\ensuremath{\mathrm{M}_{\Hd}}}

\newcommand{\mSqOneL}{\ensuremath{\mathrm{M}_{\tilde{\quark}1_\mathrm{L}}}}
\newcommand{\mSqTwoL}{\ensuremath{\mathrm{M}_{\tilde{\quark}2_\mathrm{L}}}}
\newcommand{\mSqThreeL}{\ensuremath{\mathrm{M}_{\tilde{\quark}3_\mathrm{L}}}}

\newcommand{\mSupR}{\ensuremath{\mathrm{M}_{\tilde{\uQuark}_\mathrm{R}}}}
\newcommand{\mSdownR}{\ensuremath{\mathrm{M}_{\tilde{\dQuark}_\mathrm{R}}}}
\newcommand{\mScharmR}{\ensuremath{\mathrm{M}_{\tilde{\cQuark}_\mathrm{R}}}}
\newcommand{\mSstrangeR}{\ensuremath{\mathrm{M}_{\tilde{\sQuark}_\mathrm{R}}}}
\newcommand{\mStopR}{\ensuremath{\mathrm{M}_{\tilde{\tQuark}_\mathrm{R}}}}
\newcommand{\mSbottomR}{\ensuremath{\mathrm{M}_{\tilde{\bQuark}_\mathrm{R}}}}

\newcommand{\mSlepOneL}{\ensuremath{\mathrm{M}_{\tilde{\ell}1_\mathrm{L}}}}
\newcommand{\mSlepTwoL}{\ensuremath{\mathrm{M}_{\tilde{\ell}2_\mathrm{L}}}}
\newcommand{\mSlepThreeL}{\ensuremath{\mathrm{M}_{\tilde{\ell}3_\mathrm{L}}}}

\newcommand{\mSeleR}{\ensuremath{\mathrm{M}_{\tilde{\ele}_\mathrm{R}}}}
\newcommand{\mSmuR}{\ensuremath{\mathrm{M}_{\tilde{\mu}_\mathrm{R}}}}
\newcommand{\mStauR}{\ensuremath{\mathrm{M}_{\tilde{\tau}_\mathrm{R}}}}

\newcommand{\Azero}{\ensuremath{\mathrm{A}_{\circ}}}
\newcommand{\Au}{\ensuremath{\mathrm{A}_{\uQuark}}}
\newcommand{\Ac}{\ensuremath{\mathrm{A}_{\cQuark}}}
\newcommand{\At}{\ensuremath{\mathrm{A}_{\tQuark}}}
\newcommand{\Ad}{\ensuremath{\mathrm{A}_{\dQuark}}}
\newcommand{\As}{\ensuremath{\mathrm{A}_{\sQuark}}}
\newcommand{\Ab}{\ensuremath{\mathrm{A}_{\bQuark}}}

\newcommand{\Aele}{\ensuremath{\mathrm{A}_{\ele}}}
\newcommand{\Amu}{\ensuremath{\mathrm{A}_{\mu}}}
\newcommand{\Atau}{\ensuremath{\mathrm{A}_{\tau}}}

\newcommand{\hboson}{\ensuremath{\mathrm{h}}}
\newcommand{\Hboson}{\ensuremath{\mathrm{H}}}
\newcommand{\Hpmboson}{\ensuremath{\mathrm{H}^\pm}}
\newcommand{\Aboson}{\ensuremath{\mathrm{A}^{\!\circ}}}

\newcommand{\sleptonL}{\ensuremath{\mathrm{\tilde{\ell}_\mathrm{L}}}}
\newcommand{\sleptonR}{\ensuremath{\mathrm{\tilde{\ell}_\mathrm{R}}}}

\newcommand{\squarkL}{\ensuremath{\tilde{\quark}_\mathrm{L}}}

\newcommand{\squarkR}{\ensuremath{\tilde{\quark}_\mathrm{R}}}

\newcommand{\Neutralino}{\ensuremath{\tilde{\chi}^\circ}}
\newcommand{\Chargino}{\ensuremath{\tilde{\chi}^\pm}}

\begin{document}

\begin{frontmatter}



\title{\suspect3: A \texttt{C++} Code for the Supersymmetric
and Higgs Particle Spectrum of the MSSM}

\author[a]{Jean-Lo\"{i}c Kneur}
\author[a]{Gilbert Moultaka}
\author[b]{Micha\"{e}l Ughetto}
\author[c,d]{Dirk Zerwas \corref{author}}
\author[e,f]{Abdelhak Djouadi}

\cortext[author] {Corresponding author.\textit{E-mail address:} dirk.zerwas@in2p3.fr}
\address[a]{Laboratoire Charles Coulomb (L2C), CNRS, Universit\'e de Montpellier, Montpellier; France.}
\address[b]{CPPM, Aix-Marseille Universit\'e, CNRS/IN2P3, Marseille; France.}
\address[c]{IJCLab, Universit\'e Paris-Saclay, CNRS/IN2P3, Orsay; France.}
\address[d]{also at DMLab, CNRS/IN2P3, Hamburg; Germany.}
\address[e]{CAFPE, Granada and Granada U., Theor. Phys. Astrophys., Granada; Spain.}
\address[f]{NICPB, R{\"a}vala pst. 10, 10143 Tallinn, Estonia.}

\begin{abstract}
We present the program \suspect3{} that calculates the masses and couplings of the Higgs and supersymmetric particles predicted by the Minimal Supersymmetric Standard Model (MSSM). The model is implemented in both its non-constrained version, the MSSM, and its constrained versions,  such as the minimal supergravity and the gauge or anomaly mediated supersymmetry breaking models, in which the soft supersymmetry–breaking parameters obey certain universal boundary conditions at the high energy scale.
The low energy parameters are then obtained using renormalization group equations and electroweak symmetry breaking, and all the dominant radiative corrections have been consistently implemented.  \suspect3{} is a major rewrite,  in C++ object oriented programming, of the FORTRAN code \suspect. It includes all the features of the earlier code in an improved and updated manner, and involves new  options such as compressed SUSY scenarios, an MSSM-inflation model and the possibility of using the observed Higgs mass as an input. The main features and the use of the program are explained.

\end{abstract}

\begin{keyword}
Higgs; supersymmetry; sparticle; MSSM.

\end{keyword}

\end{frontmatter}



{\bf PROGRAM SUMMARY}

\begin{small}
\noindent
{\em Program Title:} \suspect3{}                                          \\
{\em Developer's repository link:} http://suspect.in2p3.fr \\
{\em Licensing provisions(please choose one):} GPLv3  \\
{\em Programming language:} \texttt{C++}, compatible \texttt{ C++98}, \texttt{C++11}, \texttt{C++14}, \texttt{C++17}\\ 
{\em Compiler:} \texttt{gcc-4.8.5} and later (checked with \texttt{gcc-8.1.0})                                \\
{\em Nature of problem:}\\
  Supersymmetric models such as the MSSM, mSUGRA, GMSB, AMSB and others have specific parameter sets and boundary 
  conditions. \suspect3{} translates the parameter sets of the models into predictions of the Higgs and supersymmetric particles masses. The mixing matrices of the physical states as well as the mixing angles are calculated in addition
  to the scale dependent parameters.\\ 
{\em Solution method:}\\
  The spectrum of the Higgs and supersymmetric particles depends on the model, its supersymmetric parameter set and the Standard Model parameters. The evolution of the parameters as function of the energy scale is calculated by solving numerically the Renormalization Group Equations. Model dependent boundary conditions are applied at the appropriate scale. Electroweak symmetry breaking is calculated iteratively at the electroweak scale. The application of radiative corrections translate the scale dependent particle masses into the physical pole masses. \\
{\em Additional comments including restrictions and unusual features:}\\
  The parameters and physical masses are defined to be real. Warnings are issued if intermediate results are unphysical.
   \\
%
\end{small}

\section{Introduction}
\label{sec:intro}

The discovery of a Standard-Model-like Higgs boson by the ATLAS and CMS experiments~\cite{Aad:2012tfa,Chatrchyan:2012ufa} at
the LHC with a mass of about 125~GeV~\cite{Aaboud:2018wps,Sirunyan:2017exp,Aad:2015zhl} is compatible
with TeV scale supersymmetric (SUSY) theories such as
the minimal supersymmetric extension of the Standard Model (MSSM) (for reviews see e.g. ~\cite{Nilles:1983ge,Haber:1984rc,Gunion:1984yn,Gunion:1986nh,Martin:1997ns,MSSMWorkingGroup:1998fiq}).
Indeed, in the MSSM the lightest Higgs boson $h$ mass can receive large radiative corrections, as initially demonstrated in~\cite{Okada:1990gg,Ellis:1990nz,Haber:1990aw,Okada:1990vk,Ellis:1991zd}, (see e.g.~\cite{
Djouadi:2005gj,Slavich:2020zjv} for reviews and further references) and is predicted to be less than 130-140 GeV~\cite{Allanach:2004rh}.

Further motivation to study SUSY models is driven by the capability of the 
MSSM to provide a particle candidate to explain~\cite{Goldberg:1983nd,Ellis:1983ew} the relic mass--energy density of cold dark matter~\cite{Aghanim:2018eyx}.

Another motivation for supersymmetry is that it naturally stabilizes the large hierarchy between the Grand Unification scale ($\MGUT$) and the
electroweak scale ($\MEWSB$): this reduces drastically the quadratic sensitivity of radiative corrections to 
the Higgs boson masses to high scales, as was first shown in~\cite{Dimopoulos:1981zb,Sakai:1981gr}. 
Furthermore, in TeV scale SUSY models, a consistent unification of the $\rm{U}(1)_Y, \rm{SU}(2)_L$ and $\rm{SU}(3)_c$ gauge couplings
embedded in a larger symmetry group such as $\rm{SU}(5)$, is possible~\cite{Ellis:1990wk,Amaldi:1991cn,Langacker:1991an}. 

While the most general MSSM has more than 100 free parameters, more restrictive variants with only 31 real-valued parameters, 
3~gaugino masses, 15~scalar fermion masses, 9~trilinear couplings in the scalar/Higgs sector, 2 masses in the Higgs sector, $\tanb$ and the sign of $\mu$, have
been defined. Incorporating phenomenologically motivated constraints was the motivation of  
the `phenomenological~MSSM'~\cite{MSSMWorkingGroup:1998fiq}, which has only about 20~parameters.  
These parameters enter in the calculation of the SUSY particle and
Higgs boson physical masses and their couplings as well as inducing mixing between different
states. 

The number of parameters can be reduced further in well motivated
theoretical models where the soft supersymmetry--breaking (SSB) parameters obey certain
universal boundary conditions at the $\MGUT$, at a high scale $\MHIGH$ below $\MGUT$, but above $\MEWSB$, 
or at an intermediate scale $\MMESS$. 
This is for instance the case of the minimal Supergravity model~\cite{Chamseddine:1982jx,Barbieri:1982eh,Hall:1983iz}, or the gauge mediated~\cite{Dine:1993yw,Dine:1994vc,Dine:1995ag}
(see \cite{Giudice:1998bp} for a review and further references),
and anomaly mediated~\cite{Randall:1998uk,Giudice:1998xp} SUSY breaking models.

Evolving parameters between scales by solving the renormalization group equations (RGE) 
leads to an interdependence of the parameters at $\MEWSB$. Ensuring electroweak symmetry breaking (EWSB)
at $\MEWSB$ including radiative corrections further complicates the calculation.
Precise calculations of the pole masses of the Higgs bosons
and the SUSY particles involve loop contributions from most of the other (s)particles.
Therefore, the complete mass spectrum has to be known. As the boundary
conditions of the models are defined at at least two different scales, the two procedures, RGE evolution and
EWSB, have to be repeated several times to converge to a stable result.

Sophisticated spectrum calculator programs have been developed to tackle these challenges, 
such as, non-exhaustively,
\suspect~\cite{Djouadi:2002ze}, 
{\tt SOFTSUSY}~\cite{Allanach:2001kg}, {\tt SPHENO}~\cite{Porod:2003um,Porod:2011nf,Staub:2017jnp}, {\tt FeynHiggs}~\cite{Heinemeyer:1998yj,Heinemeyer:1998np,Heinemeyer:2000nz,Bahl:2018qog}, and {\tt FlexibleSUSY}~\cite{Athron:2014yba}.
\suspect3{}, a major rewrite of \suspect,
will be presented in the following. A preliminary version has been presented at a Les Houches workshop~\cite{Brooijmans:2012yi} and used e.g. in~\cite{Butter:2015fqa}.

Since its conception \suspect2{} has been updated and extended continuously leading to program that is at the same time robust and difficult to maintain. It is common, e.g., in long running experiments, to upgrade the software framework and tools to order ''organically'' grown
code and make use of new features. \texttt{C++} has several interesting features such as overloading of methods, encapsulation and inheritance which can make code maintenance and extension easier.

Spectrum calculators such as SuSpect3 tie together different fields
of physics such as collider physics and cosmology in a concrete model by providing precise predictions for the particle properties. 
As the number of degrees of freedom to be explored in such a study is large, a robust and efficient algorithm is necessary. The implementation of a new electroweak symmetry breaking variant in \suspect3{} using as input parameter the 
measured Higgs boson mass is a step to reduce the number of parameters.

\suspect3{} provides a new model linking particle physics and cosmological inflation. 
The structure of \suspect3{} facilitates its implementation which necessitated the addition of the scale 
evolution of three additional parameters and the implementation of inflation specific boundary conditions.

The paper (and user's manual) is organized as follows. The underlying physics features are discussed in \cref{sec:physics}. In \cref{sec:program} the implementation of \suspect3\ is discussed.~The main models implemented in \suspect3\ are outlined briefly in \cref{sec:modelConfig}.~A short conclusion will be given in the last section. The appendix contains information on how to install and run the program
as well as the full output of a calculation as example.

\section{Physics}
\label{sec:physics}
A basic description of the MSSM features and parameters is given in this section. For physics motivations and detailed descriptions, see the reviews of the MSSM and related models, 
e.g.~\cite{Martin:1997ns} (and references
therein).

In a SUSY theory the particle spectrum is extended with respect to the Standard Model (SM)
by associating a fermionic degree of freedom
to each bosonic degree of freedom and vice versa. Squarks $\squarkL$ and  $\squarkR$ are scalar fields associated
to the two chirality states of each quark $\quark$. In the leptonic sector,  sleptons $\sleptonR$ and 
$\sleptonL$ are scalar fields associated to each charged lepton flavor $\ell$, whereas a sneutrino is associated to the left--chiral 
neutrino of the SM. The latter is assumed to be massless for simplicity.

In the MSSM, two chiral $\rm{SU}(2)_L$ doublet Higgs superfields of opposite hypercharge
are needed to allow for supersymmetric and gauge invariant
Yukawa couplings involving up and down right-handed quark chiral superfields, as well as 
for chiral anomaly cancellation.
Their scalar components constitute
the Higgs sector. The neutral components of the latter develop vacuum expectation values, $v_u$  and $v_d$, that break 
electroweak symmetry and give masses to the up- and down-type fermions respectively. Physical states are two neutral
CP--even Higgs bosons $\hboson$ and $\Hboson$, one neutral CP--odd Higgs boson $\Aboson$ and the charged
Higgs bosons $\Hpmboson$. 

The fermionic partners of the neutral gauge and Higgs bosons mix to form the mass eigenstates, referred to 
as neutralinos, $\Neutralino_i$ with $i=1,..,4$. Similarly the charged gauge and Higgs bosons are associated
to the fermionic partners which mix to form the charginos, $\Chargino_i$ with $i=1,2$.

\subsection{MSSM Parameters}
\label{sec:params}

The description of the MSSM content follows Refs.~\cite{MSSMWorkingGroup:1998fiq,Djouadi:2002ze,Skands:2003cj}. For convenience only the parameters are defined here; for further details and the definition of the mass matrices, mixing matrices and mixing angles see~\cite{Djouadi:2002ze,Skands:2003cj}. 

The Higgs sector parameters are $\tanb$, the ratio of the vacuum expectation values $v_u/v_d$, the tree-level running mass of the pseudo scalar Higgs boson $\Aboson$ and the Higgs mass parameter $\mu$. SSB mass
parameters are $\mHu^2$ and $\mHd^2$ for the up-- and down--type Higgs fields. 

In the squark sector a SSB mass parameter associated to an 
$\rm{SU}(2)_L$ doublet is defined for 
each family: $\mSqOneL$, $\mSqTwoL$ and $\mSqThreeL$. In addition each squark flavor has 
a SSB mass parameter for its $\rm{SU}(2)_L$ singlet: 
$\mSupR$, $\mSdownR$, $\mScharmR$, $\mSstrangeR$, $\mStopR$, $\mSbottomR$. The off--diagonal entries in the squark
mass matrices depend on the SSB trilinear couplings $\Au$, $\Ad$, $\Ac$, $\As$, $\At$ and $\Ab$. Flavor mixing due 
to SSB parameters, possible in the most general model~\cite{Martin:1997ns}, 
is assumed to be absent in the following.

The structure is similar in the slepton sector. The parameters $\mSlepOneL$, $\mSlepTwoL$ and $\mSlepThreeL$ are
the SSB mass parameters associated to the $\rm{SU}(2)_L$ doublets. The SSB mass parameters associated
to the $\rm{SU}(2)_L$ singlets are $\mSeleR$, $\mSmuR$ and $\mStauR$. Three SSB trilinear couplings are defined, $\Aele$, $\Amu$ and
$\Atau$, one for each charged lepton flavor.

In the gaugino sector three SSB masses are defined, $\Mone$, $\Mtwo$ and $\Mthree$, associated respectively
to the gauge groups $\rm{U}(1)_Y, \rm{SU}(2)_L$ and $\rm{SU}(3)_c$. Without loss of generality $\Mtwo$ and $\Mthree$ are  required to be positive, while 
$\Mone$ can be either positive or negative.

\subsection{Models}
\label{sec:models}

The most general supersymmetric model implemented in \suspect3{} is the MSSM with its parameter set as defined above. 
Several variants of this model can be identified which differ only in the scale where most of the 
SUSY parameters are defined, either at $\MGUT$, $\MHIGH$ or at $\MEWSB$. A new type of SUSY model, allowing for an inflationary scenario, generically referred to as {\it Inflation} in the following, has also been implemented. We provide hereafter in \cref{sec:MSSMinflation} a short description of its main ingredients.

\subsubsection{Constrained MSSM}

The models for minimal Supergravity (mSUGRA), Gauge Mediated SUSY Breaking (GMSB) and 
Anomaly Mediated SUSY Breaking (AMSB) can be understood as subsets of the MSSM. In these models
the boundary conditions reduce the number of free parameters either at $\MGUT$ (mSUGRA, AMSB) 
or at an intermediate scale $\MMESS$ between $\MGUT$ and $\MEWSB$ (GMSB)~\footnote{See
\cref{sec:SUGRA,sec:GMSB,sec:AMSB} for details on the parameter definitions of minimal SUGRA, GMSB and AMSB  models respectively.}. In these
models $\MGUT$ may be defined as the scale of gauge coupling unification and a separate scale $\MHIGH$ is used
for the definition of the other scale--dependent parameters.

In addition to the general models GMSB and AMSB a minimal variant of these models is also provided, mGMSB and
mAMSB. In these minimal models the number of free parameters is further reduced by additional assumptions.

\subsubsection{Connecting the MSSM and Cosmological Inflation \label{sec:MSSMinflation}}

In this model \cite{PhysRevLett.97.191304}, the MSSM is extended by specific 
non-renormalizable superpotential terms lifting the so-called 
LLE or UDD MSSM flat directions \cite{Gherghetta:1995dv} that can trigger an inflationary phase.\footnote{LLE and UDD stand for the 
$\rm{SU}(3)_c\times \rm{SU}(2)_L \times \rm{U}(1)_Y$ gauge invariant operators $\epsilon_{\alpha \beta} \, L_i^\alpha L_j^\beta  \, E_k$ and $\epsilon_{a b c} \, U_i^a D_j^b  \, D_k^c$ in the superpotential, that parameterize the corresponding flat directions. $L$ denotes $\rm{SU}(2)_L$ doublets and $E,U,D$,   $\rm{SU}(2)_L$ singlets; $i,j,k$ are generation indices, $a,b,c$ color indices and $\alpha, \beta$, $\rm{SU}(2)_L$ flavor indices. The relevant cases correspond to
$i\neq j$ for LLE and $j\neq k$ for UDD.}
This involves two new {\sl a priori} free parameters:
a supersymmetric coupling $\lambda_6$ associated with a dimension-$6$ operator in the superpotential, and the corresponding SSB bi-trilinear coupling $\mathrm{A}_6$. A third important parameter is the inflaton mass $m_\phi$, which is however uniquely determined by the slepton/squark SSB masses when the inflaton rolls along the LLE/UDD
directions. $\lambda_6$ is presumably related to some unspecified UV completion of the MSSM. In contrast,
the initial condition for $\mathrm{A}_6$ can be in general linked to the other soft trilinear couplings of the MSSM at the SUSY-breaking scale, once a mediation scenario of this breaking is assumed. Following \cite{Allahverdi:2006we} we have implemented a simple relation between $\mathrm{A}_6$ and $\At$ obtained in the Polonyi model.
However, any other relation can be easily implemented.

The RGE evolution of $m_\phi$ is given by that of the  SSB
slepton/squark masses, already available in the code. Those of $\mathrm{A}_6$ and $\lambda_6$ can be extracted from the R-parity-violating MSSM (see e.g.~\cite{Allanach:2003eb}), including multiplicative factors for the non-renormalizable operators \cite{Antusch:2002ek}, and neglecting contributions
suppressed by a heavy mass scale (usually taken as $M_{planck}$) associated with these effective couplings. 
We have implemented the supplementary RGEs for all relevant LLE/UDD flavor directions including the complete gauge/gaugino
Yukawa/trilinear-soft contributions. 
The $\lambda_6$ parameter factorizes with respect to the other parameters 
and does not influence the MSSM spectrum calculation. Even so, we provide its RGE evolution alongside those of $m_\phi$ and $\mathrm{A}_6$ as the three parameters are instrumental for a precise determination of the effective potential of the inflaton at a given inflation scale. The latter is an important parameter that we implemented as an input. For a given set of MSSM parameters at the SUSY-breaking scale, $\mathrm{A}_6$ is calculated from $\At$. The ensuing knowledge of $\mathrm{A}_6$ at the inflation scale, together with the (external) cosmology constraints
(related to the determination of the Hubble flow parameters), fix uniquely $m_\phi$ and $\lambda_6$
at the inflation scale.
This cosmologically consistent inflaton mass can be compared to the running $m_\phi$ provided by \suspect3{}
at the inflation scale allowing to check the consistency of the (input) squark/slepton soft masses with cosmological
inflation. 
Incidentally, $\lambda_6$ is run to all scales to enable the check of the perturbativity of this coupling.

\subsection{EWSB}

An attractive feature of the MSSM is the radiative electroweak symmetry breaking. Essentially, the parameter
$\mHu^2$ evolved to the scale of $\MMZ$ becomes negative without leading to
tachyonic particles. As the parameter is part of the Higgs potential, the champagne bottle bottom
or Mexican hat form of the potential is radiatively generated instead of being enforced
arbitrarily.

\label{subsec:EWSB}
The minimization of the Higgs potential of the MSSM with respect to $\Hu$ and $\Hd$ leads to
a pair of quadratic equations which relate the parameters of the Higgs sector.
There are three distinct possibilities:
Either the pair $\mHu^2$, $\mHd^2$ as well as the sign of $\mu$,  
the pair $\mu$, $m_{\Aboson}$ or 
the pair $\mu$, $m_{\Aboson}^2(Q)$ can be used to calculate the other parameters consistently.

Motivated by the precise measurement of the Higgs boson mass~\cite{Aaboud:2018wps,Sirunyan:2017exp} 
EWSB can also be calculated by using the mass of the lightest neutral Higgs boson $m_{\hboson}$ as input. 
An algorithm has been recently developed in~\cite{El-Kosseifi:2022rkf}, based on diagrammatic fixed-order perturbative calculations up to two-loop precision where the trilinear coupling $\At$ becomes an output parameter. This algorithm has been implemented in \suspect3.

Requiring consistent EWSB necessitates multiple iterations to ensure that radiative corrections are calculated consistently.

\subsection{Spectrum calculation}\label{sec:spectrum}
Determining the Higgs and sparticle 
pole masses consistently with the renormalization group evolution and the EWSB mechanism poses numerous non–trivial technical problems if to be done in an accurate way, i.e. including higher order radiative corrections.
   Indeed, the latter are known to be extremely important in particular in the Higgs sector. A calculation of the various couplings is also necessary and should be performed with the same accuracy.  Note that one also has to consider radiative corrections to the Standard Model particle masses (those of the W, Z bosons, top, bottom quarks and tau lepton) and gauge and Yukawa couplings.  
  In this context, \suspect3{}, like former \suspect, follows mainly Ref.~\cite{Pierce:1996zz}, which provides most of the necessary radiative corrections  at the one-loop level for the Higgs and sparticle masses, while the leading two–loop corrections to the masses of the (neutral) Higgs bosons~\cite{Degrassi:2001yf,Brignole:2001jy,Brignole:2002bz,Dedes:2002dy,Dedes:2003km} are implemented following Ref.~\cite{Allanach:2004rh}.

\section{The program}
\label{sec:program}

In \suspect3{} the RGEs are solved to evolve  parameters from one scale to another.
Boundary conditions are defined at different scales either as input parameters or as constraints
relating different parameters to each other as function of the chosen model.
\suspect3{} ensures EWSB at $\MEWSB$ for different choices of 
input. 
Finally, radiative corrections to the Higgs and sparticle masses are implemented 
to determine precisely the pole masses, as discussed in \cref{sec:spectrum}.

The RGEs are implemented at one--loop and two--loop precision and evolve the parameters between
up to five different energy scales. At $\MGUT$, determined from the approximate 
unification scale of the gauge couplings, or at $\MHIGH$ (an arbitrary user-defined high scale), a subset of the supersymmetry breaking parameters are also unified 
depending on the models, as specified below.

In mGMSB, GMSB and Inflation models
an intermediate energy scale $\MMESS$ is defined either to implement the boundary conditions
on the SSB parameters, typically $10^{10}\,\gev$ for mGMSB, GMSB, or to calculate the 
inflaton mass. At $\MEWSB$ the minimization of the Higgs potential 
triggering EWSB is implemented. The radiative corrections
on the SM parameters are calculated at $\MMZ$ and $\tanb$ is defined at this scale.
All RGEs and radiative corrections are calculated in the $\overline{DR}$ renormalization scheme.

Due to the radiative corrections, relating RGE evolved parameters to pole masses,
and the boundary conditions being defined at at least two scales,
several iterations of the algorithm have to be performed. 
The algorithm consists of the RGE evolution between the highest scale of the model and $\MMZ$, as well as enforcing EWSB iteratively at $\MEWSB$. After convergence, the physical pole masses are calculated.

\subsection{Technical Aspects}

\suspect3{} is written in \texttt{C++} in object oriented programming. 
The main program instantiates a {\tt suspect} object. A method {\tt Run}, implementing the initialization, execution and finalization of 
the object, must be called to obtain the spectrum. {\tt Initialize, Execute} and {\tt Finalize} are public methods which could
also be called by the user. The \texttt{C++} functionality of overloading methods is useful as the same method
name can be used for different types of input. \texttt{Run}, for example, is overloaded three times, allowing to keep the same
syntax whether the input is the filename, an \texttt{SLHA4suspect} object or a pointer to the \texttt{SLHA4suspect} object.

To cope 
with different types of input, e.g., from the standard SLHA format files~\cite{Skands:2003cj} or via a {\tt C++} object, 
the {\tt Run} and {\tt Initialize} methods are overloaded. For input via an object, 
the object {\tt SLHA4suspect}, which implements SLHAio blocks in memory, has to be created, and the parameter
values in the input blocks initialized. The {\tt SLHA4suspect} object given to {\tt suspect} 
is copied internally. This has the advantage that {\tt SLHA4suspect} has to be created only once for a scan of the parameter space, the 
parameter under study can be set to a new value without reinitializing all other parameters again. 
The results of the calculation are provided as terminal output, in a file or in memory as a {\tt SLHA4suspect} object which can be retrieved 
from the {\tt suspect} object.

Each of the models discussed in \cref{sec:modelConfig} is implemented as an object. The model
construction starts from a common base class which takes care of the basic configuration. The algorithmic
flow is implemented in three generic objects, inheriting from the base class. 
These objects depend on the number of scales of the model: 
two scales with $\MEWSB$ and $\MMZ$, three scales with $\MGUT$, $\MEWSB$ and $\MMZ$, four scales with $\MGUT$, $\MMESS$, $\MEWSB$ and $\MMZ$ and five scales with $\MGUT$, $\MHIGH$, $\MMESS$, $\MEWSB$ and $\MMZ$.
The explicit models inherit from the generic model objects. The use of inheritance has lead 
to a simplification of the code by removing duplicated code copied from one model to another. The implementation
of a new model is easier.

The solution of the RGEs is coded and implemented as a separate object. 
The RGE evolution can be performed at one--loop or two--loop precision.
This enables the use of the RGE evolution independent of the full model calculation.

Several iterations between the highest scale of the model and $\MMZ$ are necessary. In the calculation
scalar SUSY mass parameters might become tachyonic at certain scales. \suspect3{} adds an error message 
to the block {\tt SPINFO} in such a case. The tachyons can be a numerical artifact induced by the coupled RGEs through
parameters, e.g., for EWSB, 
which have not converged yet or at the first iteration have not been defined yet. Therefore the errors are reset 
at each iteration at the highest scale applicable to the model.

The algorithmic flow of the determination of the EWSB output parameters is implemented in a base class. Each of the EWSB variants is implemented as a separate object, inheriting from the base class ({\tt EWSBHuHd, EWSBMAPoleMu, EWSBMA2TreeMu}). In each object the derived parameters, e.g., $m_{\Aboson}^2(Q)$ and $\mu$ in the case of {\tt EWSBHuHd}, are calculated and the boundary conditions applied.

The determination of $\At$ from the lightest Higgs boson mass is implemented as separate object {\tt EWSBMh} inheriting from the base class. Three 
objects {\tt EWSBHuHdMh, EWSBMAPoleMuMh, EWSBMA2TreeMuMh} inherit from both the {\tt EWSBMh} object and one of the objects {\tt EWSBHuHd, EWSBMAPoleMu, EWSB\-MA2Tree\-Mu}. The diamond inheritance structure due to the common {\tt EWSBBase} class is treated by virtual inheritance.
The inversion algorithm was developed initially only for \texttt{EWSBHuHd}. Inheritance made the extension of the 
other two variants simple. The implementation in \suspect2{} has not been successful.

The six variants of EWSB are complemented by a seventh variant. The non-application of EWSB consistency may also be imposed. In this mode
the derived parameters are calculated, but no iteration is performed to ensure consistency.

The particles are defined as objects which can be configured to calculate the running masses (no extra radiative
corrections), or the pole masses using the appropriate additional radiative corrections. While in
\suspect2{} variables were used both for running and pole masses, data encapsulation allows for a clean separation 
and easier maintainability.
By inheriting from a common base class an iteration over all objects is easy to implement for both variants.

After calculation of the spectrum, the absence of non-physical minima of the Higgs potential 
and the amount of fine-tuning in the EWSB conditions are checked.
A limited set of the most common precision observables is calculated as well.

\subsection{New Features and Repository}

The updates of \suspect3{} with respect to \suspect{} are:
\begin{enumerate}
    \item The first and second generation supersymmetric scalar fermion parameters, previously identical, 
    are fully independent.
    \item The full one-loop radiative corrections in the chargino and neutralino sector are calculated on 
    the chargino and neutralino  pole masses instead of applying approximate effective corrections 
    on the underlying parameters before mass diagonalization.
    \item The range of models supported by \suspect3{} has been extended.
    \item The option to use the tree-level running mass of the pseudoscalar $\Aboson$ as input for EWSB has been added.
    \item The option of using the experimentally measured Higgs boson mass as input in EWSB to determine $\At$ has been added.
    \item The input decoding has been made fully compatible with SLHA1~\cite{Skands:2003cj} with one additional block {\tt QEXTPAR} taken from SLHA2~\cite{Allanach_2009}. In particular the non-standard definition of the EWSB, i.e., setting the required EWSB in a separate input block, inherited from \suspect\ has been removed.
    \item The choice of EWSB is decoded from the defined input values instead of a required special variable.
    \item The block {\tt QEXTPAR} is decoded to determine the input scale for the $\mu$ parameter.
    \item The values of scale dependent parameters at scales other than those defined by the model is now available for all models and not only for the MSSM.
\end{enumerate}

The code is available as a gzipped tar file on the suspect website (see \ref{sec:install} for further details).
It is maintained in a gitlab repository provided by the computing center of the IN2P3. Continuous integration 
tests are implemented for all example input files provided to the user using a centos--7 image with the default gcc--g++ compiler.
The absence of memory leaks was verified for all example files with {\tt valgrind}. 

SuSpect3 was developed with the {\tt gcc} compiler version $4.8.5$. The compilation and execution were checked explicitly with \texttt{gcc-8.1.0} using \texttt{-std=c++98}, \texttt{-std=c++11}, \texttt{-std=c++14} and \texttt{-std=c++17}.

\section{Configuration of Models}
\label{sec:modelConfig}

As mentioned above, the input specifications for \suspect3{} follow closely the SLHA1 standard described in~\cite{Skands:2003cj}. 
Additionally the block {\tt QEXTPAR} of SLHA2)~\cite{Allanach_2009} is decoded to define the scale at which $\mu$ is requested. No other parameters from this block 
are used. If $m_{\Aboson}^2(Q)$ is requested, the scale for $\mu$ is used for the parameter. The models available in \suspect3\ are discussed 
briefly in this section including the inheritance.

Common to all models are six input blocks: {\tt SUSPECT\_CONF}, {\tt MODSEL}, {\tt MINPAR}, {\tt SMINPUTS}, {\tt EXTPAR} and 
{\tt QEXTPAR}.

{\tt SUSPECT\_CONFIG} defines the configuration of the calculation.
The precision of the calculation for the RGEs as well as the number of loops used can be configured. 
The accuracy of the Higgs mass and the full spectrum can also be controlled by the user:
\begin{verbatim}
BLOCK SUSPECT_CONFIG
         0     2.10000000e+01   # 21: 2-loop RGE (default), 
#11: 1-loop, 99: provided externally
         3     2.00000000e+00   # RGE accuracy: 
#1: moderate, 
#2: accurate (default), 3: expert input via index 9
         5     2.00000000e+00   # choice for sparticles 
# masses rad. corr. excluding Higgs):
# 2 ->all (recommended, default): 
# 1->no R.C. in squarks & gauginos.
         7     2.00000000e+00   # Final spectrum accuracy: 
# 1 -> 1% acc.; 2 -> 0.01 % acc.(default)
         8     2.00000000e+00   # Higgs boson masses 
# rad. corr. calculation options:
# A simple approximation (advantage=fast)  : 0
# Full one-loop calculation                : 1
# One-loop+dominant DSVZ 2-loop (default)  : 2
\end{verbatim}
If the use of an externally provided RGE set is requested via entry~0, the object in {\tt src/RgeDataAndConditionsExternal.cxx} is called. Three predefined methods have to be provided. In {\tt CalculateDerivative} the implementation of the user defined RGEs is foreseen. Its
inputs are a vector of the parameters in double precision and (optionally) a vector with boolean entries defining which
parameters should be evolved to the new scale.
The utility methods {\tt transferVector2SLHA} and {\tt transferSLHA2Vector} transfer the parameters to and from the SLHA blocks. The standard implementation 
is available in {\tt src/RgeDataAndConditions\-Internal.cxx}.

The accuracy of the solution of the RGE  equations is controlled via entry 3. Moderate precision is a relative 
precision of $1\cdot 10^{-3}$, accurate is defined as $2\cdot 10^{-5}$. Specifying the expert input, the entry is
used to specify the value of the relative precision in an allowed range from $10^{-2}$ to $10^{-6}$.

In the block {\tt SMINPUTS} the standard model parameters are defined for the input, the default values are:
\begin{verbatim}
     1     127.934     # alpha_em^(-1)(MZ) SM MSbar
     2     1.16639E-5  # G_F
     3      0.118      # alpha_s(mZ) SM, MSbar
     4     91.187      # mZ pole mass
     5      4.18       # Mb(mb) SM MSbar
     6      173.       # Mtop(pole)
     7    1.777        # Mtau(pole)
\end{verbatim}

For the scale of gauge coupling unification $\MGUT$, control is provided in the block {\tt MODSEL}: 
\begin{verbatim}
        13    -1.00000000e+00   # Gauge coupling unification scale: 
# -1: calculated, value>0: value for scale fixed
\end{verbatim}
The scale can either be calculated dynamically as part of the spectrum calculation (value $-1$) or forced to a specific input value.
This functionality was previously part of the block {\tt SUSPECT\_CONFIG}.

While in previous versions of \suspect3{} the values of the scale
dependent parameters were provided at additional scales only 
for the basic MSSM, this functionality is now provided for 
all models via the block {\tt MODSEL}:
\begin{verbatim}
        11         10   # MSSM bottom-up only: number of points
        12       1E16   # MSSM bottom-up: high scale   
\end{verbatim}
A positive value for the number of points will trigger the RGE evolution from $\MMZ$ to the scale provided at index~12. The default for this 
scale is $\MEWSB$. The steps are equidistant in $\log_{10}$.
This functionality is part of the {\tt Finalize} method implemented in the base class of the models 
and therefore inherited by all models.

In addition to the gauge coupling unification scale, \suspect{} provides furthermore control over $\MEWSB$, through index 14
of block {\tt MODSEL}:
\begin{verbatim}
        14    -1.00000000e+00   # EWSB scale: 
# -1 sqrt(stop1*stop2), value>0: value for scale fixed
\end{verbatim}
The default of $-1$ leads to the standard definition of $\MEWSB$ as the geometric mean of the stop 
quark masses as specified by SLHA~\cite{Skands:2003cj}. 
Specifying a positive value at this index overrides this choice and fixes $\MEWSB$ to be the chosen value 
as suggested e.g. in~\cite{AguilarSaavedra:2005pw}. 

After these common settings, the selection of the specific model via the {\tt MODSEL} block will now be described. 
For some of the models, example input files are provided with the distribution 
in the subdirectory {\tt examples}, see \ref{sec:install} for download and installation. In the following
the relative pathname of the input file will be mentioned where available.

\subsection{MSSM}
\label{sec:MSSM}

The {\it MSSM} is configured in the {\tt MODSEL} block as: 
\begin{verbatim}
         1          0   # MSSM:0, mSUGRA:1,....
\end{verbatim}

The parameter $\tanb$, defined at $\MMZ$, as well as the sign of $\mu$ are configured in the block {\tt MINPAR}:
\begin{verbatim}
    3     2.0E+01  # tanbeta(MZ)
    4      1.0     # sign(mu)
\end{verbatim}

The other parameters are defined in the block {\tt EXTPAR}:
\begin{verbatim}
BLOCK EXTPAR
         1     5.25225703e+02   # M_1
         2     9.54406026e+02   # M_2
         3     2.54213069e+03   # M_3
        11    -3.11637739e+03   # A_t
        12    -5.31256168e+03   # A_b
        13    -3.41866265e+03   # A_tau
        14    -4.60623573e+03   # A_u
        15    -5.91979687e+03   # A_d
        16    -3.53443255e+03   # A_e
        17    -4.60623573e+03   # A_c
        18    -5.91979687e+03   # A_s
        19    -3.53443255e+03   # A_mu
        21     4.86427234e+05   # M^2_Hd
        22    -4.01017724e+06   # M^2_Hu
#        23     2.01330021e+03   # mu(EWSB)
#        24     4.77432382e+06   # m^2_A_run(EWSB)
#        26     2.14083063e+03   #  mA
        31     1.18999153e+03   # M_eL
        32     1.18999153e+03   # M_muL
        33     1.12125121e+03   # M_tauL
        34     1.00009204e+03   # M_eR
        35     1.00009204e+03   # M_muR
        36     8.24069856e+02   # M_tauR
        41     2.44756888e+03   # M_q1L
        42     2.44756888e+03   # M_q2L
        43     2.02265804e+03   # M_q3L
        44     2.35965026e+03   # M_uR
        45     2.35965026e+03   # M_cR
        46     1.47723188e+03   # M_tR
        47     2.34881332e+03   # M_dR
        48     2.34881332e+03   # M_sR
        49     2.24171006e+03   # M_bR
\end{verbatim}
Following the SLHA standard, as 
index~0 is not specified, the scale of the parameters is $\MEWSB$. To use a fixed user chosen $\MEWSB$, as discussed above, the parameter at 
index~14 in the {\tt MODSEL} block has to be set to the chosen value.

The example shown here is for EWSB defined by $\mHu^2$, $\mHd^2$ and sign of $\mu$.
For EWSB with $\mu$,$m_{\Aboson}$ the indices 21 and 22 must be replaced by the indices 23 and 26. 
If EWSB with $\mu$,$m_{\Aboson}^2(Q)$ is required, indices 23 and 24 must be provided.
The sign of $\mu$ specified in {\tt MINPAR} is ignored in these two cases as $\mu$ in {\tt EXTPAR} is a 
signed parameter.

The detection of the EWSB option is done automatically: the default is defined as $\mHu^2$,$\mHd^2$, then {\tt EXTPAR}
is checked for the alternative inputs which override the default. No configuration variable has to be set in {\tt SUSPECT\_CONFIG}
for these three possibilities. \suspect{} checks the consistency of the input, i.e., if the input is ambiguous, e.g., both the running and 
the pole mass of the $\Aboson$ are given, the execution is stopped and an error
message is printed.

Using the Higgs boson mass as input parameter instead of $\At$ requires further specifications.
In a large part of the supersymmetric parameter space the determination of $\At$ from the Higgs
boson mass has a four-fold ambiguity. Each one of the solutions, ordered in $\At$ is requested
via the block {\tt SUSPECT\_CONFIG}:
\begin{verbatim}
    4     1.01000000e+02   # 100-103: EWSB with Mh four solutions
\end{verbatim}
This type of EWSB has been tested only for the benchmark point discussed in~\cite{El-Kosseifi:2022rkf}.

The request for this type of EWSB is triggered by adding 
$m_{\hboson}$ in block {\tt SMINPUTS}:
\begin{verbatim}
        25     1.25012052e+02   # h
\end{verbatim}
If the Higgs boson mass is specified in the input, the value of $\At$ in the {\tt EXTERNAL} block 
is ignored.

In case partial unification of SSB parameters is requested, the corresponding
unified parameter should be provided in {\tt MINPAR} and the parameters removed from {\tt EXTPAR}, e.g. \cref{sec:SUGRA,sec:LSMSSM,sec:HSMSSM}.

The mixing between the right-- and left--handed sfermions due to the trilinear couplings of the first two generations 
is negligible due to the smallness of the corresponding SM fermion mass. The trilinear couplings are nevertheless part
of the model specification since they have an impact on the RGE evolution of the parameters, 
especially between $\MGUT$ and $\MEWSB$, thus modifying the predicted spectrum. The mixing matrix for the smuons is
calculated for the prediction of the supersymmetric contributions to the anomalous magnetic moment of the
muon.

Under the umbrella of the {\it MSSM} several distinct variants 
can be identified which are discussed in the following.

\subsubsection{Low Scale MSSM}
\label{sec:LSMSSM}

For the input configured, as discussed above, only two scales are relevant, $\MEWSB$ and $\MMZ$. The model, 
{\it Low Scale MSSM}, inherits
in this case from the generic model with two scales. For SSB parameters not specified
in {\tt EXTPAR} the corresponding parameter in {\tt MINPAR} will be used as defined in~\cite{Skands:2003cj}. This
enables the user to easily define partially unified models.

As {\tt EXTPAR(0)} is not given, the scale $\MEWSB$ is calculated 
as geometric mean of the stop pole masses. 
For convenience
the scale is listed as output in block {\tt MODSEL}:
\begin{verbatim}
        15     1.72300647e+03   # OUTPUT ONLY: EWSB scale
\end{verbatim}

An example input is provided in {\tt examples/LowScaleMSSM.in}. 

\subsubsection{High Scale MSSM}
\label{sec:HSMSSM}

Another variant of the MSSM is calculated if in the block {\tt EXTPAR} the scale parameter is given:
\begin{verbatim}
         0     1.34489071e+16   #  scale for params in EXTPAR 
\end{verbatim}
If the value is $-1$, the model described in \cref{sec:LSMSSM} is calculated. If the scale is defined ($>0$), the scale
is interpreted as $\MHIGH$ from which the scale dependent 
parameters have to be evolved to $\MEWSB$ before the calculation
of the physical masses. As in \cref{sec:LSMSSM} partially unified models can be defined easily by
specifying the corresponding {\tt MINPAR} parameter and not specifying the {\tt EXTPAR} parameters.

The model inherits from the
generic model with four scales: $\MGUT$ for gauge coupling unification, $\MHIGH$ which in this case is the SUSY-breaking scale, $\MEWSB$ and $\MMZ$.

An example input is provided in {\tt examples/\allowbreak HighScaleMSSM.in}.  

To ensure the equality of the gauge coupling unification scale $\MGUT$ and the 
input scale for the supersymmetric parameters $\MHIGH$, index~0 of {\tt EXTPAR} and index~13 of {MODSEL}
have to be set to the same value. In this case the GUT scale is imposed by the user. To obtain 
the equality of the two scales and calculate dynamically $\MGUT$ the
model configuration described in \cref{sec:SUGRA} should be used.

The pMSSM~\cite{MSSMWorkingGroup:1998fiq} can be configured by specifying the same SSB model
parameters for the first and second generation, but separate values for the quark and leptonic sectors. 
This cannot be done through the block {\tt MINPAR} by omitting the parameters of one generation as only
a single scalar SSB mass is defined.

\subsection{Minimal Supergravity}
\label{sec:mSUGRA}
The model {\tt minimal Supergravity} is requested via {\tt MODSEL}:
\begin{verbatim}
         1        1   # MSSM:0, mSUGRA:1,....
\end{verbatim}
For a review of minimal supergravity models see e.g. \cite{Martin:1997ns} and original references therein.
A common parameter $\mzero$ is used for all scalar supersymmetry breaking parameters. The gaugino
mass parameters are unified as $\mhalf$. A common trilinear parameter $\Azero$ initializes
all individual trilinear parameters.  
The three parameters $\mzero$, $\mhalf$ and $\Azero$, defined at $\MGUT$, are input in the {\tt MINPAR} block:
\begin{verbatim}
     1    900.    # m0
     2   1200.    # m1%2
     5  -3000.    # A0
\end{verbatim}
in addition to the sign of $\mu$ and $\tanb$. This model inherits from the generic three scales model as in this case $\MGUT$ and $\MHIGH$ are identical.

An example input is provided in {\tt examples/mSUGRA.in} .

\subsubsection{Supergravity inspired Models}
\label{sec:SUGRA}

A variant of mSUGRA is also implemented. If the 
{\tt EXTPAR} parameters are given, the SSB parameters given in {\tt MINPAR} are 
overwritten. This can be limited to a single parameter in {\tt EXTPAR}. An example input is provided in {\tt examples/SUGRA.in}.

The resulting model is identical to the model described in \cref{sec:HSMSSM} if $\MHIGH$ is specified at index~0 of {\tt EXTPAR}. 
However if $\MHIGH$ is not specified, the scales 
$\MGUT$ and $\MHIGH$ are identical and the gauge coupling unification, either dynamical (default) or user imposed through 
index~13 of {\tt MODSEL}, determines both scales. Having a dynamically calculated gauge unification scale exactly equal
to the SUSY-breaking scale cannot be achieved with the standard SLHA MSSM input in contrast to this model.

\subsection{Gauge Mediated SUSY Breaking}
\label{sec:GMSB}

The {\tt Gauge Mediated SUSY-Breaking} model, for a review see \cite{Giudice:1998bp}, 
is defined via {\tt MODSEL} as:
\begin{verbatim}
         1       102   # MSSM:0, mSUGRA:1,...
\end{verbatim}
Its input parameters are defined in the block {\tt MINPAR}:
\begin{verbatim}
     1    100E3  # Lambda_susy
     2    200E3  # Lambda_mess
    52    1      # GMSB: N messenger SU2
    53    1      # GMSB: N messenger SU3
\end{verbatim}
in addition to the sign of $\mu$ and $\tanb$. The number of U(1) messengers is calculated from
the number of SU(2) and SU(3) messengers as defined in Eq.~(24) of~\cite{Djouadi:2002ze}. 
The boundary conditions are calculated and applied at $\MMESS$ (index 2).

The GMSB model inherits from the generic model with four scales. 
An example input is provided in {\tt examples/GMSB.in}.

\subsection{Minimal Gauge Mediated SUSY Breaking}
\label{sec:mGMSB}

The {\tt minimal Gauge Mediated SUSY-Breaking} model is defined via {\tt MODSEL} as:
\begin{verbatim}
         1        2   # MSSM:0, mSUGRA:1,...
\end{verbatim}
The input in block {\tt MINPAR} is simplified with respect to GMSB in \cref{sec:GMSB} as
the number of messengers SU(2) and SU(3) is fixed to unity.

mGMSB inherits from the GMSB model. 
An example input is provided in {\tt examples/mGMSB.in}.

\subsection{Anomaly Mediated SUSY Breaking}
\label{sec:AMSB}

The {\it Anomaly Mediated SUSY Breaking} model~\cite{Randall:1998uk,Giudice:1998xp} is defined via {\tt MODSEL}:
\begin{verbatim}
         1      103   # MSSM:0, mSUGRA:1,...
\end{verbatim}
Its input parameters are specified in {\tt MINPAR} in addition to the sign of $\mu$ and $\tanb$ are:
\begin{verbatim}
    1    450.    # m0
    2    60E3    # M_3%2 gravitino mass
    5    3.      # c squarkLeft doublet 
    6   -1.      # c up type squarkR singlet
    7   -1.      # c down type squarkR singlet
    8    1.      # c sleptonLeft doublet
    9    1.      # c sleptonRight singlet
   10   -2.      # c Higgs u-type
   11   -2.      # c Higgs d-type
\end{verbatim}
The boundary conditions are calculated and applied at $\MGUT$. The input values with the indices $5-11$ are 
model dependent factors which multiply the square of the common scalar mass parameter in the calculation of the boundary
conditions at $\MGUT$.

AMSB inherits from the generic four scales model as the two scales $\MGUT$ and $\MHIGH$ can be separated via {\tt MODSEL}
by requesting a fixed gauge coupling unification scale.

\subsection{Minimal Anomaly Mediated SUSY Breaking}
\label{sec:mAMSB}

The {\tt minimal Anomaly Mediated SUSY-Breaking} model is defined via {\tt MODSEL} as:
\begin{verbatim}
         1        3   # MSSM:0, mSUGRA:1,...
\end{verbatim}
The input of {\tt MINPAR} is simplified with respect to \cref{sec:AMSB} as the multiplicative factors
for the scalar masses (indices $5-11$) are set to unity.

The mAMSB model inherits from AMSB. 
An example input is provided in {\tt examples/mAMSB.in}.

\subsection{Compressed SuSy}
\label{sec:CompSusy}

The {\tt Compressed SUSY} model~\cite{Martin:2007gf} is requested in the {\tt MODSEL} block as: 
\begin{verbatim}
          1        50   # MSSM:0, mSUGRA:1,
\end{verbatim}
The gaugino mass parameters are calculated from a common parameter and three
arbitrary coefficients $C_i$ of the respective representations of the symmetric product of two adjoint representation 
of $SU(5)$ (i.e. ${\bf 24}$). For $C_i\ne 0$ the gaugino mass degeneracy at GUT scale is lifted.

The input is defined in block {\tt MINPAR} as:
\begin{verbatim}
     1    100.    # m0
     2    250.    # m1%2
     5    -100.   # A0
     3    1.0e+01 # tanbeta(MZ)
     4    1.0     # sign(mu)
     6    0.22    #C24
     7    0.00    #C75
     8	  0.00    #C200
\end{verbatim}
The model inherits from the generic four scale model. An example input is provided in {\tt examples/CompressedSuSy.in} 

\subsection{Inflation}
\label{sec:Inflation}

The {\tt Inflation} model is defined via {\tt MODSEL} as:
\begin{verbatim}
         1      200   # MSSM:0, mSUGRA:1,...
\end{verbatim}
The inflation specific configuration is in the block {\tt SUSPECT\_CONFIG}:
\begin{verbatim}
        10     1.00000000e+00   # Type of Inflation LLE=1, UDD=2
        11     1.00000000e+00   # index of L   or U
        12     2.00000000e+00   # index of  L  or  D
        13     3.00000000e+00   # index of   E or   D
        14     1.00000000e-02   # lambdaInflation(Inflation)
        \end{verbatim}
Either LLE or UDD can be chosen for the definition of the inflaton mass.
The generational indices (values of indices $11-13$ for $i,j,k$) can be chosen to specify a flat direction. The consistency of the choices noted in \cref{sec:MSSMinflation} is checked. Inconsistent choices result in an error and stop of the calculation. 
The value of $\lambda_6$ (index $14$) at the inflation scale may also be specified as input. If the value is not 
specified, 1 is used as 
value.

The input parameters are specified in the block {\tt MINPAR}, e.g.,
\begin{verbatim}
         2     8.50000000e+14   #  INFLATION: Inflation scale 
\end{verbatim}
for the inflation scale in addition to $\mu$ and $\tanb$. In {\tt EXTPAR} the parameters for the different variants of the MSSM described in \cref{sec:MSSM} are specified.

Compared to the MSSM the Inflation model has two additional parameters, $A_6$ and $\lambda_6$. 
As boundary condition at the SUSY-breaking scale, $A_6$ is calculated from $\At$. The parameter $\lambda_6$ has no 
influence on the calculation of the spectrum.

The two additional parameters as well as the inflaton mass are output in a scale dependent block {\tt INFLATION}:
\begin{verbatim}
BLOCK INFLATION Q=8.50000000e+14
         1     4.33371532e+04   # A6
         2     1.14084300e+00   # lambda6
         3     6.87081350e+03   # mPhi
\end{verbatim}
where the inflaton mass is the square root of the mean of the soft breaking squared masses.

The model inherits from the generic five scale model: $\MGUT$ for gauge coupling unification, $\MHIGH$ for the supersymmetry breaking parameters, $\MMESS$ the inflation scale, $\MEWSB$ and
$\MMZ$.

\subsection{External}

An {\tt external} model can be provided. Specifying {\tt MODSEL} with
\begin{verbatim}
         1      999   # MSSM:0, mSUGRA:1,...
\end{verbatim}
calls this model. It can be provided by implementing the object {\tt ModelExternal} inheriting from {\tt ModelBase}
and compile against the \suspect3{} library.

\section{Calculation}
\label{sec:benchmark}

As an example the calculation of a mSUGRA parameter
set is performed for the parameters:
\[
\begin{array}{lcl}
\mzero     & = & 900\gev \\
\mhalf     & = & 1200\gev \\
\Azero     & = & -3000\gev \\
\tanb      & = & 20 \\
\mu        & = & +1 \\
\end{array}
\]
The full output is provided with the software distribution in the subdirectory {\tt examples} in the file 
{mSUGRA.out}, only some of the blocks
are listed here (and comments were edited):
\begin{verbatim}
BLOCK MODSEL
         1     1.00000000e+00   # MSSM:0, mSUGRA:1, mGMSB=2, 
# mAMSB=3, (+100 for nonminimal GMSB, AMSB), 
# Inflation: 200, External: 999
         3     0.00000000e+00   # MSSM particle content:0, 
# External: 99
        13    -1.00000000e+00   # Gauge coupling unification 
# scale: -1: calculated, value>0: value for scale fixed
        14    -1.00000000e+00   # EWSB scale: 
# -1: sqrt(stop1*stop2), 
# value>0: value for scale fixed
BLOCK SUSPECT_CONFIG
         0     2.10000000e+01   # 21: 2-loop RGE (default), 
# 11: 1-loop, 99: provided externally
         3     2.00000000e+00   # RGE accuracy: 1: moderate, 
# 2: accurate (default), 3: expert input via index 9
         5     2.00000000e+00   # choice for sparticles 
# masses rad. corr. excluding Higgs:
#               2 ->all (recommended, default); 
# 1->no R.C. in squarks & gauginos.
         7     2.00000000e+00   # Final spectrum accuracy: 
# 1 -> 1% acc.; 2 -> 0.01 % acc.(default)
         8     2.00000000e+00   # Higgs boson masses 
# rad. corr. calculation options: 
# A simple (but very good) approximation (advantage=fast)  : 0
# Full one-loop calculation                                : 1
# One-loop  + dominant DSVZ 2-loop (default,recommended)   : 2
BLOCK MINPAR
         1     9.00000000e+02   #  MSUGRA: m0 
         2     1.20000000e+03   #  MSUGRA: m_1/2 
         3     2.00000000e+01   #  tanbeta(mZ)
         4     1.00000000e+00   #  sign(mu)
         5    -3.00000000e+03   #  MSUGRA: A0 
BLOCK SMINPUTS
         1     1.27934000e+02   # alpha_em^-1(M_Z)^MSbar
         2     1.16639000e-05   # G_F [GeV^-2]
         3     1.18000000e-01   # alpha_S(M_Z)^MSbar
         4     9.11870000e+01   # M_Z pole mass
         5     4.18000000e+00   # mb(mb)^MSbar
         6     1.73000000e+02   # mt pole mass
         7     1.77700000e+00   # mtau pole mass
        13     1.05658360e-01   # muon pole mass
        24     1.42000000e+00   # charm pole mass
BLOCK HMIX Q=1.72400304e+03
         1     2.01330599e+03   # mu(Q)
         2     1.92663632e+01   # tanbeta(Q)
         3     2.44003070e+02   # vev(Q)
         4     4.77434506e+06   # MA^2(Q)
BLOCK GAUGE Q=1.34489088e+16
         1     5.44086472e-01   # gprime(Q) DRbar
         2     7.02412615e-01   # g(Q) DRbar
         3     6.98524506e-01   # g_3(Q) DRbar
BLOCK MSOFT Q=1.72400304e+03
         1     5.25225703e+02   # M_1
         2     9.54406026e+02   # M_2
         3     2.54213069e+03   # M_3
        21     4.86427234e+05   # M^2_Hd
        22    -4.01017724e+06   # M^2_Hu
        31     1.18999153e+03   # M_eL
        32     1.18999153e+03   # M_muL
        33     1.12125121e+03   # M_tauL
        34     1.00009204e+03   # M_eR
        35     1.00009204e+03   # M_muR
        36     8.24069856e+02   # M_tauR
        41     2.44756888e+03   # M_q1L
        42     2.44756888e+03   # M_q2L
        43     2.02265804e+03   # M_q3L
        44     2.35965026e+03   # M_uR
        45     2.35965026e+03   # M_cR
        46     1.47723188e+03   # M_tR
        47     2.34881332e+03   # M_dR
        48     2.34881332e+03   # M_sR
        49     2.24171006e+03   # M_bR
BLOCK MSOFT Q=1.34489088e+16
         1     1.20000000e+03   # M_1
         2     1.20000000e+03   # M_2
         3     1.20000000e+03   # M_3
        21     8.10000000e+05   # M^2_Hd
        22     8.10000000e+05   # M^2_Hu
        31     9.00000000e+02   # M_eL
        32     9.00000000e+02   # M_muL
        33     9.00000000e+02   # M_tauL
        34     9.00000000e+02   # M_eR
        35     9.00000000e+02   # M_muR
        36     9.00000000e+02   # M_tauR
        41     9.00000000e+02   # M_q1L
        42     9.00000000e+02   # M_q2L
        43     9.00000000e+02   # M_q3L
        44     9.00000000e+02   # M_uR
        45     9.00000000e+02   # M_cR
        46     9.00000000e+02   # M_tR
        47     9.00000000e+02   # M_dR
        48     9.00000000e+02   # M_sR
        49     9.00000000e+02   # M_bR
BLOCK AU Q=1.72400304e+03
  1  1    -4.60623573e+03   # A_u(Q) DRbar
  2  2    -4.60623573e+03   # A_c(Q) DRbar
  3  3    -3.11637739e+03   # A_t(Q) DRbar
BLOCK ALPHA
  -5.20508237e-02   # Mixing angle in the Higgs sector
BLOCK MASS
         1     5.00000000e-03   # d quark mass
         2     1.50000000e-02   # u quark mass
         3     1.90000000e-01   # s quark mass
         4     1.42000000e+00   # c pole mass
         5     4.81354684e+00   # b pole mass from mb(mb)_MSbar
         6     1.73000000e+02   # t pole mass
        11     5.11000000e-04   # electron mass
        13     1.05658360e-01   # muon mass
        15     1.77700000e+00   # tau mass
        23     9.11870000e+01   # Z
        24     8.04878483e+01   # W+
        25     1.24911649e+02   # h
        35     2.14067721e+03   # H
        36     2.14080999e+03   # A
        37     2.14258711e+03   # H+
   1000001     2.51767568e+03   # ~d_L
   1000002     2.51629841e+03   # ~u_L
   1000003     2.51767568e+03   # ~s_L
   1000004     2.51629841e+03   # ~c_L
   1000005     2.06541208e+03   # ~b_1
   1000006     1.49367951e+03   # ~t_1
   1000011     1.20014656e+03   # ~e_L
   1000012     1.19637831e+03   # ~nu_eL
   1000013     1.20016602e+03   # ~mu_L
   1000014     1.19637811e+03   # ~nu_muL
   1000015     8.18162868e+02   # ~tau_1
   1000016     1.12587444e+03   # ~nu_tauL
   1000021     2.59653958e+03   # ~g
   1000022     5.20861806e+02   # ~chi_10
   1000023     9.85073613e+02   # ~chi_20
   1000024     9.85246261e+02   # ~chi_1+
   1000025    -2.00867791e+03   # ~chi_30
   1000035     2.01085133e+03   # ~chi_40
   1000037     2.01181065e+03   # ~chi_2+
   2000001     2.41345033e+03   # ~d_R
   2000002     2.42394519e+03   # ~u_R
   2000003     2.41345033e+03   # ~s_R
   2000004     2.42394519e+03   # ~c_R
   2000005     2.30563356e+03   # ~b_2
   2000006     2.09869582e+03   # ~t_2
   2000011     1.00344813e+03   # ~e_R
   2000013     1.00342496e+03   # ~mu_R
   2000015     1.13292099e+03   # ~tau_2
\end{verbatim}
At $\MGUT$ the gauge couplings are unified exactly for the couplings $g_1$ and $g_2$ when taking into account
the factor $\sqrt{\frac{3}{5}}$ from SU(5). They unify approximately with $g_3$.
$\MEWSB$ as geometric mean of the two stop masses is about $1.7\tev$. At $\MEWSB$ $\tanb$ is slightly 
smaller than its input value which is defined at $\MMZ$.

The supersymmetry soft breaking masses, unified
at $\MGUT$, after RGE evolution lead to a non-unified spectrum.
The benchmark point was chosen to lead to a mass of the lightest neutral Higgs boson
of $\sim 125\gev$. 

\subsection{Comparisons}

\suspect2{} has been compared extensively with other spectrum generators in the past, e.g.~\cite{Allanach:2003jw}. Furthermore a comparison with the spectrum calculator SoftSUSY was performed in~\cite{Lafaye:2007vs} showing 
an excellent agreement within the theoretical uncertainties.

The implementation of the algorithms in \suspect3{} is similar, but not identical to the one in \suspect2{}, 
therefore differences are expected, though these must be less than the theoretical uncertainties of the calculation. During the development of \suspect3{} extensive comparisons have been performed with \suspect2{} for each step of the calculation to ensure
that each calculation, given the same input, results in identical results at machine level precision.
For several specific parameter sets the comparisons were performed with extreme error settings, such as increasing the number of iterations
in the calculation of the pole mass for the Higgs sector, leading to excellent agreement. 

In the wino region the lightest neutralino and the lightest chargino are degenerate
in their running masses at the EWSB scale. When calculating the radiative corrections in order to predict the pole masses, the mass of the 
chargino is shifted upward by $\mathcal{O}(100)$~MeV. Such a parameter set is very sensitive to the details of 
the implementation of radiative corrections. Detailed comparisons between \suspect3{} and \suspect2{} showed an excellent agreement.

\begin{figure}[htb]
    \centering
    \includegraphics[width=0.49\textwidth]{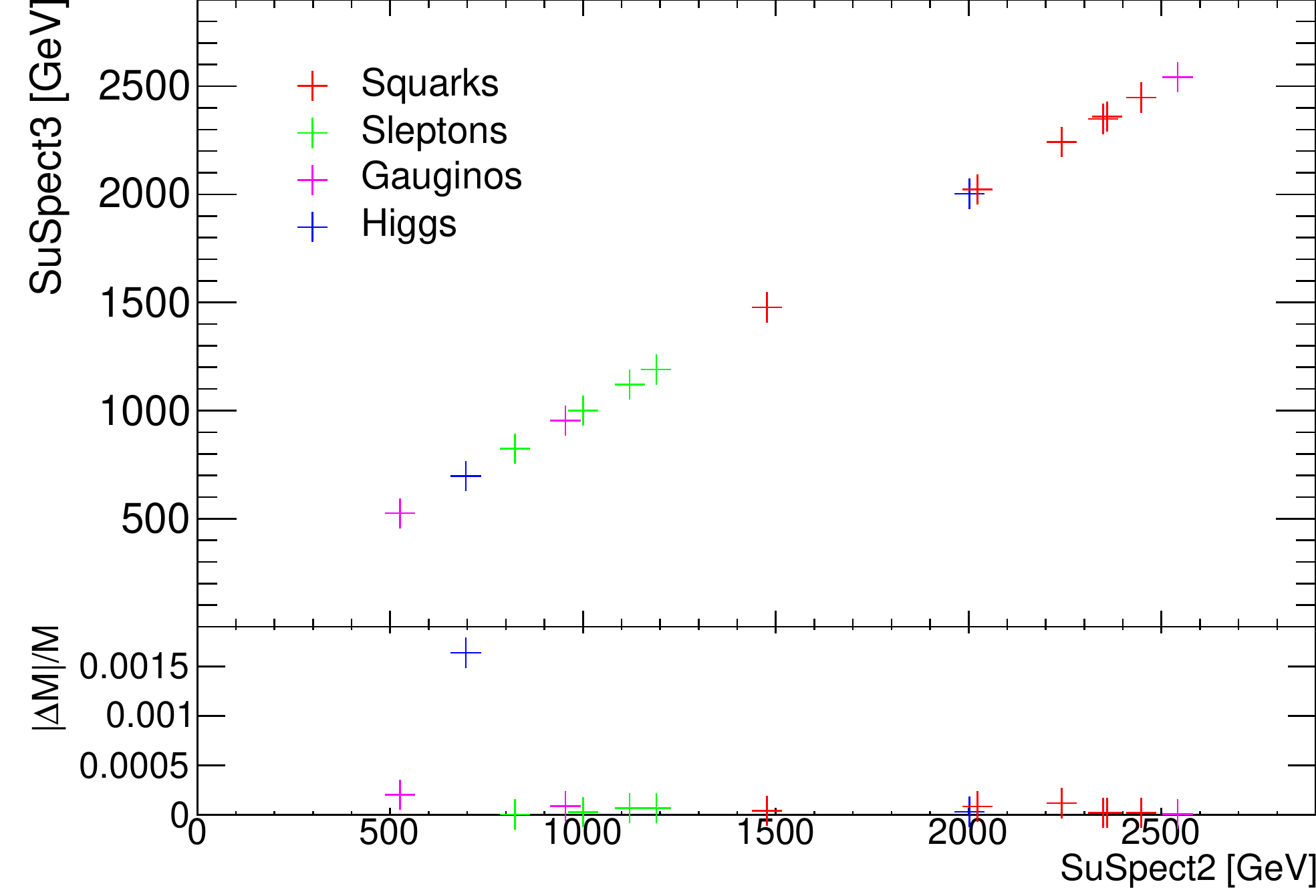}
    \includegraphics[width=0.49\textwidth]{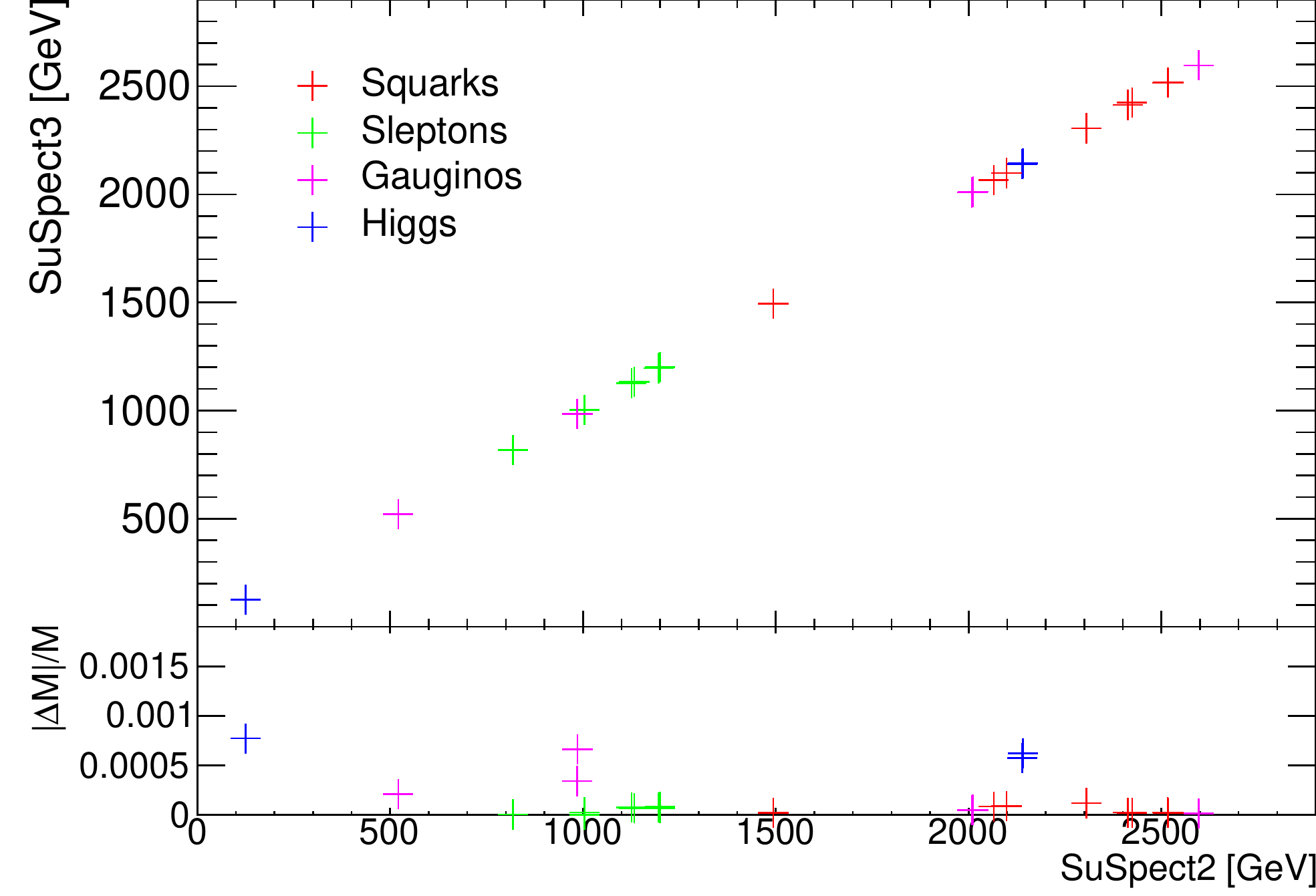}
    \caption{(Left) The supersymmetry breaking masses of the benchmark point \cref{sec:benchmark} at the EWSB scale are compared using \suspect2{} and \suspect3{}. On the lower panel the normalized relative difference is shown. (Right) The pole masses are shown for \suspect3{} as function of the \suspect2{} calculation.
    On the lower panel the normalized relative mass difference.}
    \label{fig:S2S3}
\end{figure}

In \cref{fig:S2S3} an illustration for 
the mSUGRA point defined and discussed in \cref{sec:benchmark} is shown for the default precision settings. 
On the left the RGE evolution
is tested by comparing
the running supersymmetry soft breaking masses at the EWSB scale between \suspect3{} and \suspect2{}. 
In the Higgs sector the absolute value 
is displayed. The agreement is for most masses better than $10^{-4}$. The yukawa couplings on the bottom mass
agree to the per mil level explaining the slightly worse agreement for $\mHd$.
On the right panel
the pole masses are compared. The agreement is better than per mil for all masses and for most masses at the level of $5\cdot 10^{-5}$.

The mSUGRA point has also been compared to the result of a \texttt{SPHENO} calculation. For this comparison the EWSB scale 
in \suspect3{} was set to 1~TeV as \texttt{SPHENO} was run with the SPA convention~\cite{AguilarSaavedra:2005pw}. The value
of $\tan\beta(\MMZ)$ in \suspect3{} was adjusted so that $\tan\beta(\MEWSB)=20$ as output by \texttt{SPHENO}. The running
mass of the block \texttt{MSOFT} and the pole masses of block \texttt{MASS} agree at the level of several $10^{-3}$ for most 
of the masses.

\section{Conclusion}

The MSSM is an attractive extension of the Standard Model providing a light Higgs boson, gauge coupling 
unification and a candidate for dark matter. \suspect3, written in \texttt{C++}, is a sophisticated tool
translating input parameters into a particle spectrum of Higgs bosons and supersymmetric particles. 

The new structure of \suspect3{}, a major rewrite of \suspect, allows for an easier extension of the 
supported models.~The RGE evolution of new parameters and the addition of new variants of
EWSB is simplified. \suspect3{} provides a model for cosmological inflation as well as
the RGE evolution of its parameters. The new variant of EWSB, where the measured Higgs boson mass 
is used as input instead of $\At$, written initially for a single EWSB variant, is another example of 
the advantage of the new structure.

Superymmetry remains an attractive model to tie together
different fields in a concrete model valid up to \MGUT. The exploration 
of multi dimensional parameter space depends on precise predictions and efficient algorithms. 
Future work will center on the improvement of the new EWSB variant to reduce the number of dimensions of these
explorations.

\section*{Acknowledgements}
We would like to thank the members of the GDR Supersymmetry and the GDR/IRN Terascale for numerous discussions over the years.
We are grateful to Jan Heisig, Inga Str\"umke and Amit Adhikary for feedback on the earlier versions of \suspect3{}.
A.D. is supported by the ERC grant MOBTT86 and by the Junta de Andalucia through the Talentia Senior program as well as by the grants A-FQM-211-UGR18, P18-FR-4314 with ERDF and PID2021-128396NB-I00.
J.-L.K. and G.M. have received partial support from the European Union’s Horizon 2020 research and innovation programme under the Marie Skłodowska-Curie grant agreement No 860881-HIDDeN.


\begin{appendix}

\section{Installing SUSPECT3}
\label{sec:install}

The procedure to install \suspect3{} and run a mSUGRA calculation from an input file shipped with the package is:
\begin{verbatim}
wget http://suspect.in2p3.fr/tar/suspect3.tar.gz
mkdir myDir
mv suspect3.tar.gz myDir
cd myDir
tar xvfz suspect3.tar.gz
./configure
make
suspect3 -d examples/mSUGRA.in
\end{verbatim}

In case {\tt wget} is not available:
\begin{verbatim}
Open an internet browser
enter the address: http://suspect.in2p3.fr
download suspect3.tar.gz
\end{verbatim}

\end{appendix}



\bibliographystyle{elsarticle-num}
\bibliography{suspect3}







\end{document}